# Realization of Insulating Buffer Layers via MOCVD-Grown Nitrogen-Doped (010) β-Ga$_2$O$_3$

Rachel Kahler,[1*,a)] Carl Peterson,[1*] and Sriram Krishnamoorthy[1,a)]

[1]*Materials Department, University of California Santa Barbara, Santa Barbara, California, 93106, USA*

*Rachel Kahler and Carl Peterson contributed equally to this work*

___________________________________

[a)] **Author to whom correspondence should be addressed.  Electronic mail:** rachelkahler@ucsb.edu and sriramkrishnamoorthy@ucsb.edu

We present MOCVD-grown, nitrogen-doped β-Ga$_2$O$_3$ films as an insulating buffer layer on Fe-doped (010) β-Ga$_2$O$_3$ substrates in lieu of 49 % HF treatment to remove unintentional silicon at the substrate-epitaxial layer growth interface. N-doped layer thickness and NH$_3$ flow were systematically varied to experimentally determine the lowest nitrogen concentration and thickness of the buffer layer needed to fully compensate the interfacial silicon peak. The NH$_3$ molar flow rate was varied from 200 sccm to 1800 sccm. Results showed fully insulating N-doped layers for samples with NH$_3$ flow rates ≥ 1200 sccm and a thickness of 50 nm. This study demonstrates the efficacy of in-situ, controllably doped nitrogen buffer layers as a mitigation method for unintentional interfacial silicon at the substrate-epitaxial layer growth interface.

Energy efficient devices continue to be of high importance as global energy consumption increases. Wide bandgap semiconductors can create energy efficient devices by accessing high electric fields, which enables compact power devices. SiC and GaN have already demonstrated the advantages of wide bandgap semiconductors in the electronics market today[1], and β-$Ga_2O_3$ can further take advantage of these characteristics with a wider bandgap of 4.6 eV, a theoretical breakdown field of 8 MV/$cm^2$[2], and mobilities up to 200 $cm^2$/Vs [3]. Its high Baliga's Figure of Merit, greater than that of SiC and GaN, leads to an improved relationship between on resistance and breakdown field, with the availability of shallow donors[4]. Additionally, large area $Ga_2O_3$ boules can be melt grown, providing a scalable platform for gallium oxide power electronics[5].

Previous reports[6–8] have shown the presence of an unintentional silicon peak at the β-$Ga_2O_3$ substrate-homoepitaxial layer interfaces. The concentration is theorized to originate from ambient concentration of siloxanes[9,10] or from chemicals used in substrate polishing[11]. This peak is detrimental to lateral transistor performance because it creates a low mobility electron channel farther from the gate terminal[12]. This parasitic channel adds to the RC delay and hinders the fast switching properties of the device, which is detrimental to high frequency performance. The unintentional silicon peak also leads to a higher pinch-off voltage and increased subthreshold slope, which can be non-uniform across a wafer, negatively impacting device uniformity and performance[13]. Additionally, this peak can cause increased leakage and premature breakdown[6]. Research has shown that the silicon peak can be removed with a 49% HF dip[6], but this process is time sensitive, and the peak will return in approximately 10 minutes post-dip under ambient exposure[14].

In lieu of using 49% HF, the detrimental electron channel can be mitigated using an insulating buffer to compensate the interfacial charge. Beyond just charge compensation, an insulating buffer is helpful for high performance vertical and lateral FETs[15–17], and selective area doping and regrowth in both lateral and vertical devices[18]. An insulating buffer can be created using an acceptor such as nitrogen, magnesium, or iron[17–23]. N has a deep acceptor level of 2.9 eV below the conduction band[23], which means that there is a lesser chance of inadvertent ionization than other acceptors such as iron, which has a level of 0.86 eV below the conduction band[22,24]. Magnesium, another acceptor, has been experimentally determined to have a level of 0.65 eV above the valence band[25], but has been seen to have severe memory effects when used in MOCVD growth. Magnesium diffusion has a strong dependence on temperature, leading to unwanted concentrations of the element in subsequently grown layers[26,27]. Ion implantation can be used, but this can lead to additional diffusion issues when performing post-implant annealing[18]. N-doping, on the other hand, has less significant memory and surface riding or diffusion effects in MOCVD and can also be done in-situ to avoid the time sensitive HF cleaning process.

N-doping can be performed using various sources of nitrogen in MOCVD[17,21]. $N_2O$ has been explored for N doping beyond its typical use as an alternative oxygen source in MOCVD growth. But, it has an incorporation efficiency that is very dependent on temperature and pressure, and is kinetically inert below 700 °C and 75 Torr[12,21]. In contrast, nitrogen doping using $NH_3$ allows for controllable N-doping with only flow rate variation[21]. Since $NH_3$ has a decomposition temperature of approximately 600 °C[28], it allows for deposition at standard $Ga_2O_3$ growth temperatures.

Films were grown using an Agnitron Agilis 100 vertical cold-wall reactor with a remote injection showerhead distance of 18 cm. TEGa was used as the gallium precursor, $O_2$ as the oxygen source, $SiH_4$ as the silicon source, and argon as the carrier gas. $NH_3$ was the nitrogen source, and was diluted to 50 ppm with $N_2$ balance gas. All samples were grown at a chamber pressure of 60 Torr. Before growth, the samples were cleaned in acetone, methanol, and DI water, but the 49% HF dip[6] was intentionally left out to investigate if the insulating N-doped layer compensated the Si peak. The N-doped layer was grown at 910 °C, as higher temperature growth has shown less hydrogen incorporation in the films[21]. H incorporation is predicted to cause compensation of N in the film and act as a shallow donor[29–31]. Additionally, it has been experimentally seen that increased H concentration in the films causes a rougher surface[32,33]. Then, the ammonia flow was shut off, and a 150 nm Si-doped channel layer with an intended doping of $5 \times 10^{17}$ cm$^{-3}$ was grown at 810 °C (Fig. 1(a)).

Too much nitrogen in the films could detrimentally affect the crystal quality of the buffer layer and increase the number of unintentional impurities in the silicon doped channel if there is surface segregation or diffusion of nitrogen into the active channel layer. However, the concentration of N in the buffer layer should be sufficient to compensate unintentional silicon impurities at the epitaxial layer-substrate interface. Thus, a systematic experiment was performed to find the minimum $NH_3$ flow (N incorporation) and insulating layer thickness required. The initial N-doped layers were grown at a constant thickness of 300 nm while varying $NH_3$ flows from 200-1800 sccm to find the minimum $NH_3$ flow needed to compensate the silicon peak. Devices were fabricated with C-V pads and buffer leakage structures. All structures were mesa isolated using a $BCl_3$ inductively coupled plasma (ICP) or reactive ion etch (RIE) etch. C-V pads were fabricated using Ni/Au Schottky contacts with the lift-off method. Buffer leakage structures were lithographically defined and metal was deposited using a liftoff method. The contacts were then annealed in $N_2$ at 470 °C to ensure Ohmic characteristics.

Capacitance-voltage characteristics (C-V) were measured on each sample with varying NH$_3$ flow to determine if the silicon peak at the substrate-epitaxy interface was compensated, and the net charge concentration was calculated. As shown in Fig. 1(c), the 200 sccm flow rate showed no compensation as the carrier concentration plot shows an increase in charge density at the substrate-epitaxy interface. A similar increase in charge density is seen at the substrate-epitaxy interface for the 400 sccm flow. The 800 sccm sample shows partial compensation, with some carriers seen at the interface. Both the 1200 and 1800 sccm NH$_3$ flow samples see no carrier spike at the substrate-epitaxy interface, meaning that the concentration of nitrogen in the buffer layer is higher than that of the silicon at the interface, assuming nitrogen is a compensating acceptor[23]. To further verify full compensation, buffer leakage test structures were fabricated. These structures were ohmic test structures with 5-25 μm contact spacing, mesa isolated to the N-doped layer so that the resistance of the buffer layer could be examined. The N-doped layers were compared to a rectangular test structure (TLM) that contacted the conductive Si-doped channel layer. As is seen in Fig. 1(d), the 200 sccm sample had the same resistance as the reference conductive contacts, labeled as Ref. In contrast, the 400 and 800 sccm samples showed higher resistance, indicating partial compensation of the silicon peak at the interface. The 1200 and 1800 sccm samples showed even higher resistance, indicating a very resistive buffer layer, which is desired for lateral field effect transistors. These buffer leakage test structures corroborate the successful compensation of regrowth interface charge observed in the C-V measurements (Fig. 1(b) and (c)).

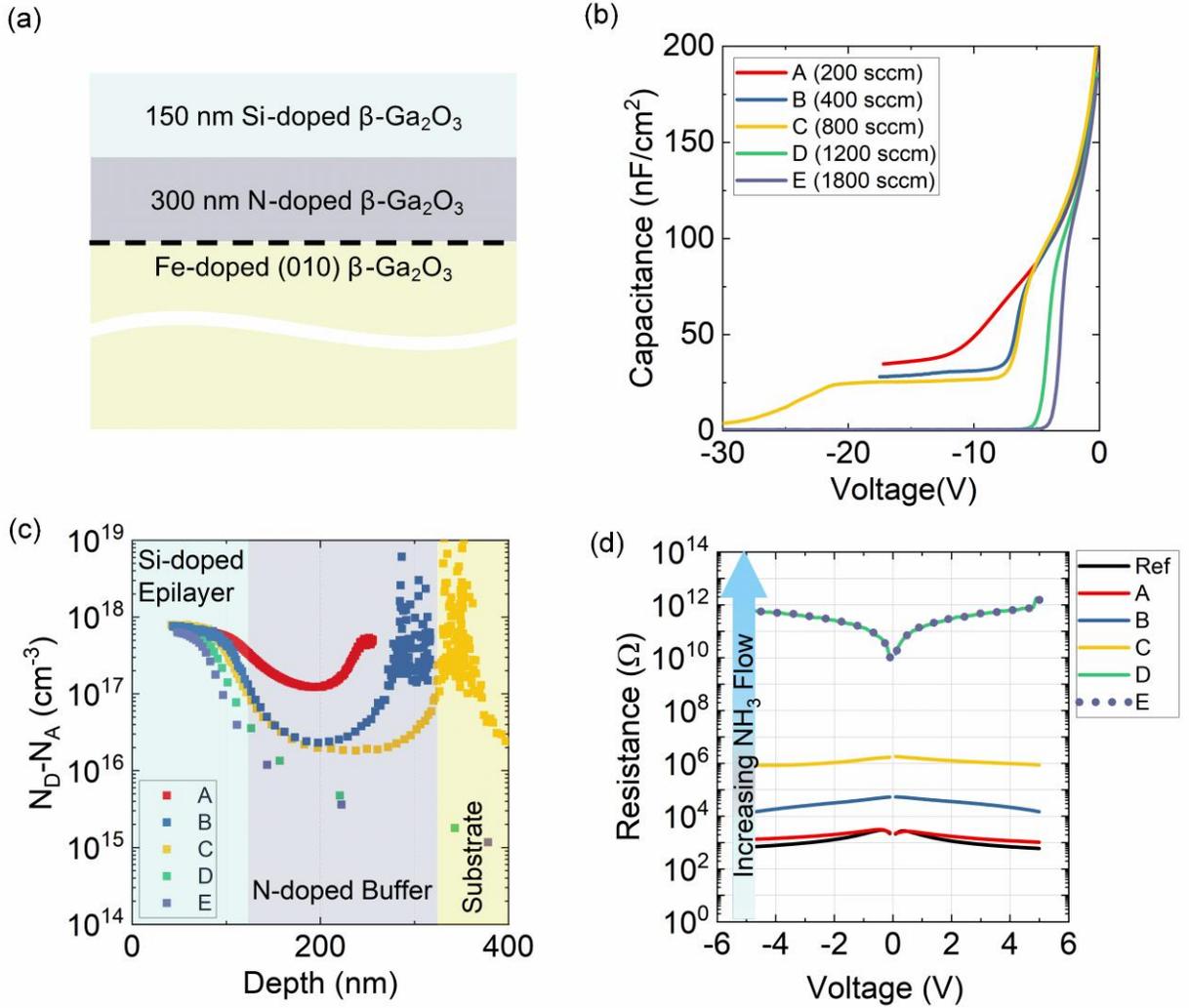

FIG. 1. (a) Epitaxial structure of Si-doped channel layer with N-doped buffer (b) C-V and (c) charge density profile of samples with $NH_3$ flows ranging from 200-1800 sccm (d) Buffer leakage compensation data as a function of $NH_3$ flow rate, where Ref is an ohmic pad contacting the Si-doped channel.

To quantify the amount of N in the films, a SIMS stack (Fig. 2(a)) was grown with 100 nm thick N-doped $\beta$-$Ga_2O_3$ buffer layers with flows of 200, 1200, and 1800 sccm interspaced with 150 nm UID layers. SIMS performed by EAG Eurofins (Fig. 2(b)) showed a N concentration of 1.7 x $10^{19}$ atoms/$cm^3$ in the 1200 and 1800 sccm layers, respectively. In the 200 sccm layer, the concentration of N was 1.5 x $10^{18}$ atoms/$cm^3$. The Si peak at the substrate-epitaxy interface had a concentration of 6.6 x $10^{18}$ atoms/$cm^3$. For a layer to be fully compensating, it must have an acceptor concentration greater than the silicon concentration at the interface. This is clearly seen in the 1200 sccm layer, which had a higher concentration of N than Si. This confirmed what was observed in C-V measurements in Fig. 1(c), where the 1200 sccm sample had no charge at the interface, while the 200 sccm sample had a large amount of charge at the interface. The H concentration, which was consistently higher than the detection limit of $5 \times 10^{16}$ cm$^{-3}$ across the entire SIMS profile, follows the increase in the N concentration in the N-doped layers. The profiles of the nitrogen in the N-doped buffer layers do not perfectly resemble the "top hat" that would be seen without diffusion or surface riding of N atoms into the UID layers separating the N-doped layers. This shows that there was considerable surface riding or diffusion of N into the other layers. Diffusion should be minimal into the other layers, considering that N is not seen to diffuse at or below 1100 °C[18], and all layers were grown significantly below this temperature condition. But, the profile of the N doped layers is symmetric into both the layers before and after it, showing that surface riding is most likely not the culprit in these films. Further experiments are required to quantify if the N diffused into other grown layers.

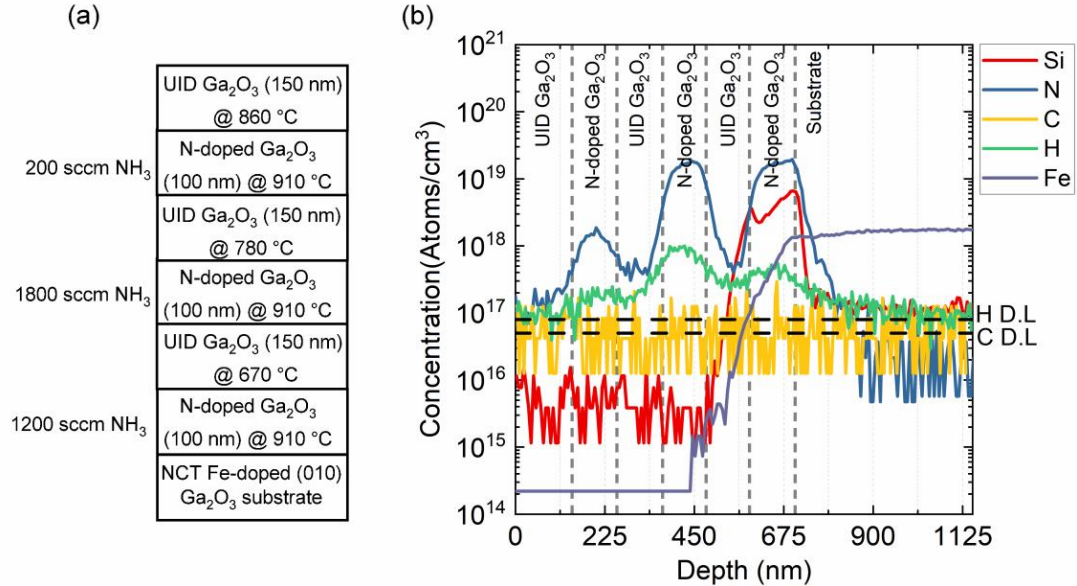

FIG. 2. (a) Schematic of MOCVD-grown SIMS stack with N-doped layers grown at 1200, 1800 and 200 sccm NH$_3$ flows, and UID layers grown at 670, 780, and 860 °C and (b) SIMS data showing concentrations of Si, N, C, H, and Fe. C concentration is at the detection limit for the entire scan.

After finding the minimum compensating NH$_3$ flow of 1200 sccm for a 300 nm buffer layer, the thickness of the buffer layer was optimized. Three samples were grown with thicknesses of 10, 50, and 100 nm at an ammonia flow of 1200 sccm. C-V pads, buffer leakage test structures, and TLMs to the Si-doped channel layer were fabricated on the samples. C-V data from the samples (Fig. 3(a) and (b)) showed fully compensating behavior for all thicknesses, meaning that no charge peak was observed at the substrate-epitaxy interface. But, for the 10 nm thick sample, C-V data from different areas of the sample showed non-uniform charge density, indicating potential issues with compensation as seen in Fig. 3(c). These uniformity issues were not observed in the 50 and 100 nm thick samples. Buffer leakage data from these samples in Fig. 3(d) showed highly resistive behavior in all three samples, indicating compensating buffers. Thus, the optimized buffer was determined to be 50 nm thick with a 1200 sccm NH$_3$ flow.

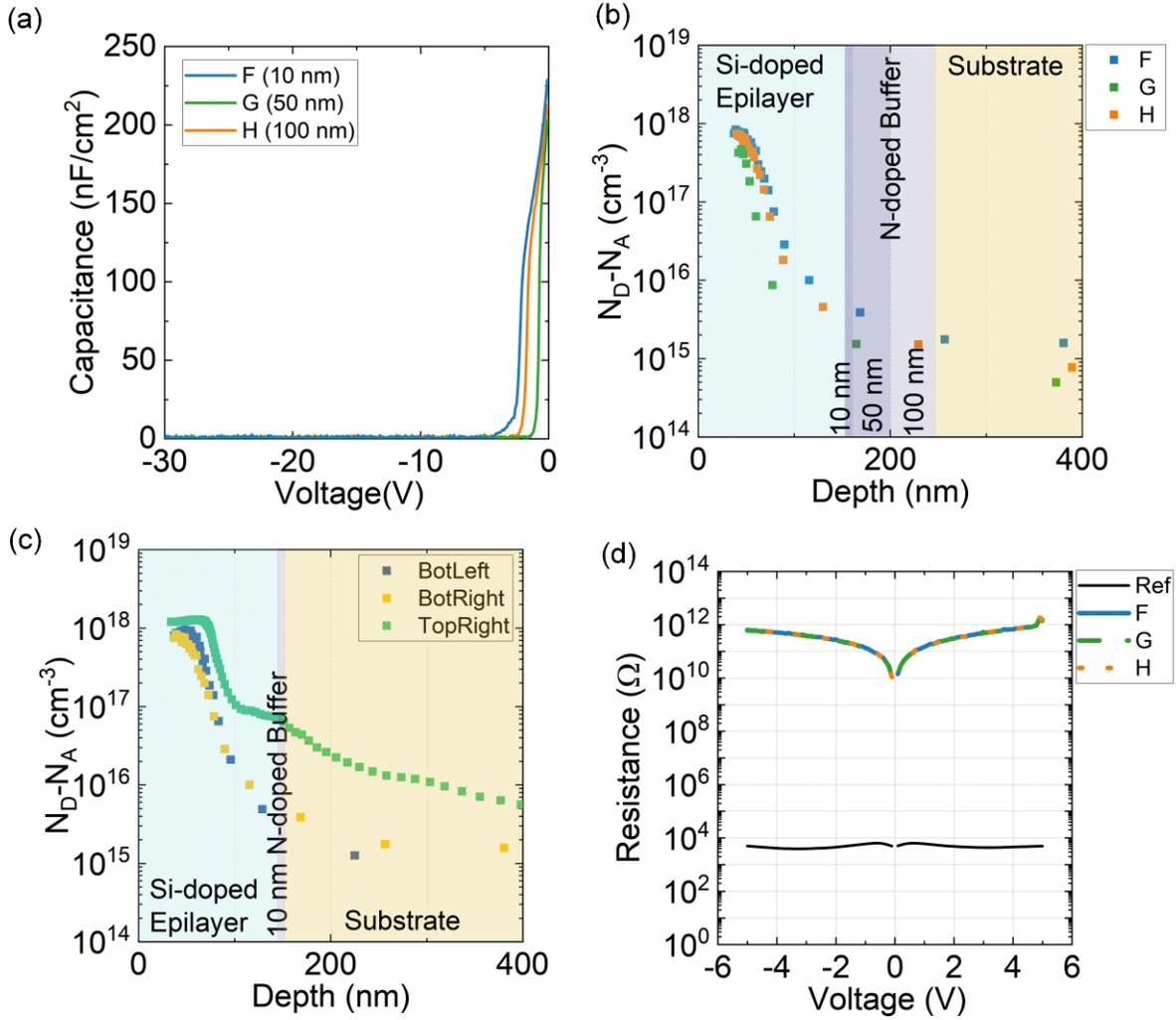

FIG. 3. (a) C-V and (b) Charge density profiles of N-doped buffer samples with varying thicknesses (c) Non-uniform charge density in 10 nm thick N-doped buffer samples (d) Buffer leakage data from all 3 samples.

AFM scans of all grown samples at varying flows and thicknesses showed 3.85-13.8 nm RMS roughness, with the surface roughness increasing as N doped buffer layer thickness increased as is seen in Fig. 4(a). In contrast, there was no observed trend between roughness and $NH_3$ flow. The morphology of the sample was comparable to what is typically seen for TEGa films grown on (010) $Ga_2O_3$ substrates, with grooves extending along the [001] direction due to enhanced adatom diffusion lengths (Fig. 4(b))[3]. The increased roughening with a thicker growth can potentially be explained by increased H incorporation with thicker buffer layers, which causes roughening in the films[32,33]. SIMS data showed increased H concentration in the 1200 and 1800 sccm layers of $5 \times 10^{17}$ and $9.3 \times 10^{17}$ atoms/cm$^3$, respectively, and a H concentration of 1-3 $\times 10^{17}$ atoms/cm$^3$ in the UID layers, which is above the detection limit.

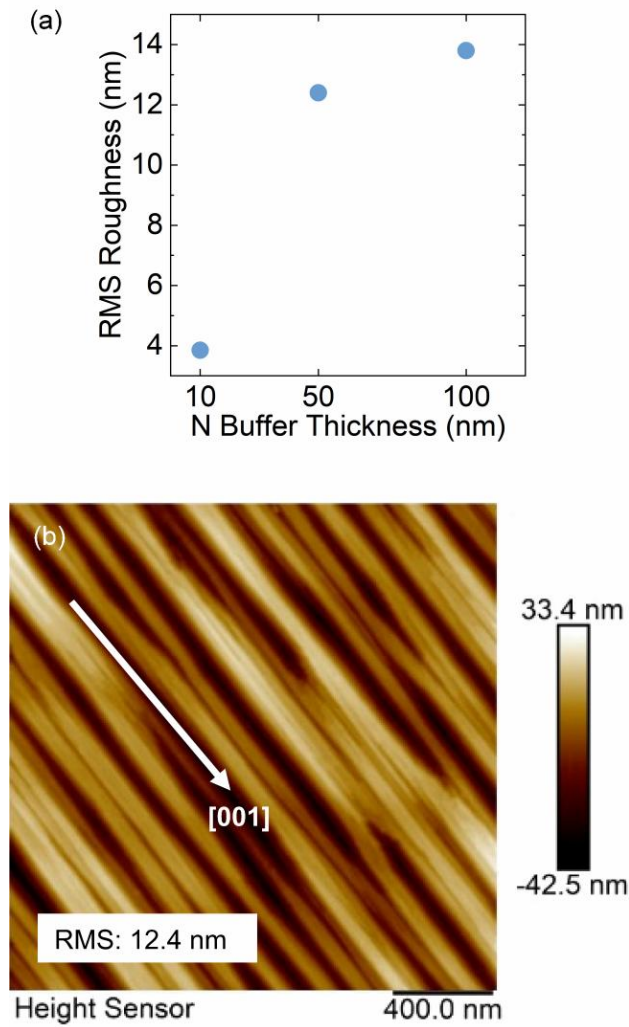

FIG. 4. (a) RMS roughness from AFM scans of samples F, G, and H and (b) AFM scan of sample F, the 50 nm N buffer sample.

Integrated channel charge trends were also explored in the samples. With increasing amounts of N, the integrated channel charge decreases, as is seen in Fig. 5. This could be due to backside depletion from the N-doped buffer layer and N incorporation into the Si-doped channel via surface riding or diffusion. The integrated channel charge was compared for the 800, 1200, and 1800 sccm flow samples with a 300 nm N buffer. Increased $NH_3$ flow from 800 sccm to 1800 sccm decreased the channel charge by approximately 2 times, from $1.1 \times 10^{13}$ to $5.49 \times 10^{12}$ $cm^{-2}$.

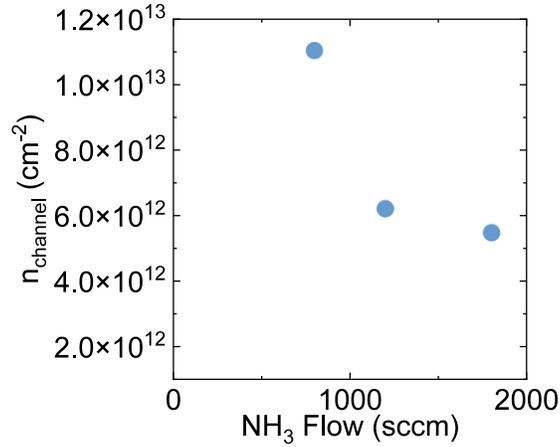

FIG. 5. Integrated channel charge from 300 nm N buffers with varying $NH_3$ flow rates.

The observed channel charge depletion was explored using a simple one-dimensional electrostatics model. By treating the Si-doped channel and N-doped layer junction as a metal-semiconductor junction with the Fermi level pinned at the acceptor level in the N-doped layer, the depletion width from the N-doped substrate into the Si-doped channel can be calculated. The depletion width obtained using this model was 72 nm, which showed that there was significant back-depletion from the N-doped layer into the Si-doped channel, decreasing the channel charge. This corroborated what was seen from integrated channel charge values from C-V measurements in Fig. 5. It was theorized that improved channel charge could potentially be observed by the insertion of a UID layer between the N-doped layer and the Si-doped channel, as this would allow the back-depletion to occur into the UID layer, not the Si-doped channel, and improve the channel charge.

An insulating N-doped buffer layer was developed for mitigation of the interfacial silicon peak seen at the substrate-epitaxy interface. The optimal buffer was 50 nm thick with a dilute $NH_3$ flow of 1200 sccm. The ability to mitigate parasitic charge layers in devices without the need for time sensitive processes such as HF treatment allows for efficient devices to be fabricated, and N doped layers hold immense promise for use in both vertical and lateral devices as an insulating layer. More work is needed to be done to further improve channel charge, such as the addition of thicker UID layers between the N-doped buffer and channel layer. This work demonstrates N doped layers as an effective insulating layer.

## ACKNOWLEDGMENTS

The authors acknowledge funding from the ARPA-E ULTRAFAST program (DE-AR0001824), Coherent / II-VI Foundation Block Gift Program, and the DoD SMART Scholarship for Service Program. A portion of this work was performed at the UCSB Nanofabrication Facility, an open access laboratory.

## AUTHOR DECLARATIONS

The authors have no conflicts to disclose.

## DATA AVAILABILITY

The data that support the findings of this study are available from the corresponding author upon reasonable request.